\documentstyle[fleqn,epsf,graphicx]{article}
\begin{document}
\title{Some Properties of the Spin-Polarized Attractive Hubbard Model}
\author{A. Kujawa and R. Micnas}
\date{}

\maketitle

\vspace{-1.1cm}
\begin{center}
\begin{footnotesize}
\emph{Solid State Theory Division, Faculty of Physics,\\ Adam Mickiewicz University,\\ Umultowska 85,
61-614 Pozna\'n, Poland}
\end{footnotesize}
\end{center}

\begin{center}
\begin{footnotesize}
\end{footnotesize}
\end{center}

\begin{abstract}
 The influence of a Zeeman magnetic field on the superconducting characteristics of the attractive Hubbard model was investigated. The ground state and temperature phase diagrams were obtained for a fixed number of particles. Two critical magnetic fields were found for the first order phase transition from the superconducting to the normal state for $n\neq 1$. For some range of parameters a reentrant transition was found and gapless superconductivity was obtained. The significance of the Hartree term was also analyzed.

\end{abstract}

\section{Introduction}
The impact of the Zeeman coupling between the spins of the electrons and an applied magnetic field on superconductivity has been analyzed for many years  \cite{C}-\cite{gruenberg}. One of the well-known results concerning this influence is the discovery of the \emph{so-called} Clogston limit. In the weak coupling regime ($s$-wave pairing), at $T=0$, the superconductivity is destroyed through the paramagnetic effect and there is a first-order phase transition to the normal state at a universal value of the critical magnetic field $h_{c}=\Delta_0 \slash \sqrt{2} \approx 0.707\Delta_{0}$ \cite{C}, where $\Delta_{0}$ is the gap at $T=0$, $h=0$. Further investigations have revealed that in the presence of magnetic field the formation of Cooper pairs across the spin-split Fermi surface with non-zero total momentum ($\vec{k} \uparrow$, $-\vec{k}+\vec{q} \downarrow$) is possible, leading to the \emph{so-called} Fulde and Ferrell \cite{fulde}, and Larkin and Ovchinnikov \cite{larkin} (FFLO) state, which is favored (against the normal state) up to $h_c^{FFLO}=0.754\Delta_0$. Because of the very strong destroying influence of the orbital effect on the superconductivity, it is hard to observe the above-mentioned effects in ordinary superconductors. However, the development of experimental techniques in cold atomic Fermi gases allows investigation of the spin polarization on superfluidity \cite{daniel}-\cite{Ketterle} in a simple way. Recent experiments \cite{Ketterle} have indicated that in the density profiles of trapped Fermi mixtures with population imbalance there is an unpolarized superfluid core in the center of the trap and a polarized normal state surrounding this core (phase separation). 

In this paper we analyze the influence of the Zeeman term on the superfluid characteristics of a lattice fermion (the spin-polarized attractive ($U<0$) Hubbard) model \cite{MicnasModern}:
\begin{equation}
\label{ham}
H=\sum_{ij\sigma} (t_{ij}^{\sigma}-\mu \delta_{ij})c_{i\sigma}^{\dag}c_{j\sigma}+U\sum_{i} n_{i\uparrow}n_{i\downarrow}-h\sum_{i}(n_{i\uparrow}-n_{i\downarrow}),
\end{equation}
where: $\sigma = \uparrow ,\downarrow$, $n_{i\sigma}=c_{i\sigma}^{\dag} c_{i\sigma}$,  
$t_{ij}^{\sigma}$ -- hopping integral, $U$ -- on-site interaction, $\mu$ -- chemical potential. The Zeeman term can be created by an external magnetic field (in ($g \mu_B \slash 2$) units) or by a spin population imbalance in the context of cold atomic Fermi gases.

We consider the case of pairing only with $\vec{q}=0$. The gap parameter is defined by: $\Delta=-\frac{U}{N}\sum_i \langle c_{i \downarrow} c_{i \uparrow} \rangle =-\frac{U}{N}\sum_{\vec{k}} \langle c_{-\vec{k} \downarrow} c_{\vec{k} \uparrow} \rangle$. 
Applying the broken symmetry Hartree approximation, we obtain the grand canonical potential $\Omega$ \cite{kujawa} and the free energy $F$:
\begin{eqnarray}
\label{pot}
&&\hspace{-0.5cm}\frac{F}{N}=\frac{1}{4} Un(2-n)-\mu (1-n)+\frac{1}{4}UM^2 -\frac{|\Delta|^2}{U}\\
&&\hspace{-0.5cm}-\frac{1}{\beta N}\sum_{\vec{k}} \ln \Bigg (2\cosh\frac{\beta (E_{\vec{k}\uparrow}+E_{\vec{k}\downarrow})}{2}+2\cosh\frac{\beta (-E_{\vec{k}\uparrow}+E_{\vec{k}\downarrow})}{2}\Bigg),\nonumber
\end{eqnarray}
where: $N$ -- number of the lattice sites, $\beta=1/k_B T$, $M=n_{\uparrow}-n_{\downarrow}$ -- spin magnetization (polarization), $n_{\sigma}=\frac{1}{N} \sum_{\vec{k}} \langle c_{\vec{k} \sigma}^{\dag} c_{\vec{k} \sigma} \rangle$ -- spin-up and spin-down electron density, $n=n_{\uparrow}+n_{\downarrow}$ -- electron concentration, 
$E_{\vec{k}\downarrow}= (-t^{\downarrow}+t^{\uparrow})\Theta_{\vec{k}}+ \frac{UM}{2}+h+\omega_{\vec{k}}$, $E_{\vec{k}\uparrow}=(-t^{\uparrow}+t^{\downarrow})\Theta_{\vec{k}}-\frac{UM}{2}-h+\omega_{\vec{k}}$, $\omega_{\vec{k}}=\sqrt{((-t^{\uparrow}-t^{\downarrow})\Theta_{\vec{k}}-\bar{\mu})^2+|\Delta|^2}$ are quasiparticle energies, $\bar{\mu}=\mu-\frac{Un}{2}$. Here $\Theta_{\vec{k}}=\sum_{l=1}^{d} cos(k_l a_l)$ ($d=3$ for three-dimensional lattice), $a_l=1$ in further considerations.
Using (\ref{pot}), one can easily get the equations for the gap, particle number (determining $\mu$) and magnetization \cite{kujawa}: $\frac{\partial F}{\partial \Delta}=0$, $\frac{\partial F}{\partial \mu}=0$, $M=-\frac{1}{N}\frac{\partial F}{\partial h}$, respectively. In the above equations, we have taken into account the spin dependent Hartree term. This method can be called a BCS-Stoner approach. In the following, we set $t^{\uparrow}=t^{\downarrow}=t$ and use $t$ as the unit. 

\section{Results}
We have performed an analysis of the influence of magnetic field on superfluidity, focusing on the ground state and the temperature phase diagrams for a fixed electron concentration and indicating differences with respect to the case of a fixed chemical potential \cite{kujawa}. Only selected results will be presented below, while an extended paper will be published elsewhere \cite{A}.

Fig. 1 shows the ground state phase diagrams for a simple cubic lattice, i.e. critical magnetic fields for the first order superconducting-normal (SC-NO) phase transition vs. the chemical potential (a) and the particle concentration (b), for three values of the on-site attraction, ranging from a weak to intermediate coupling. If the number of particles is fixed and $n\neq 1$, one obtains two critical magnetic fields in the phase diagrams \cite{kujawa}. The two critical fields (Fig. 1b) define the phase separation (PS) region between the superconducting phase with the particle density $n_s$ and the normal state with the density of particles $n_n$, as opposed to the case of a fixed chemical potential (Fig. 1a). Moreover, the increasing on-site attraction widens the range of occurrence of PS. One can also distinguish the partially polarized ($P=(n_{\uparrow}-n_{\downarrow})\slash (n_{\uparrow} + n_{\downarrow})<1$) normal state (NO-I) and the fully polarized ($P=1$) normal state (NO-II) in the phase diagrams. The same regions appear for $n>1$ because of the particle-hole symmetry. The first order transition lines at $T=0$ have been determined numerically from the condition $\Omega_s^{T=0} =\Omega_n^{T=0}$ (where $\Omega_n^{T=0}$ and $\Omega_s^{T=0}$ denote the grand canonical potential of the normal ($\Delta=0$, $P\neq 0$) and the superconducting ($\Delta \neq 0$, $P=0$) state, respectively.). 

\begin{figure}[]
\begin{center}
\includegraphics[width=0.34\textwidth,angle=270]{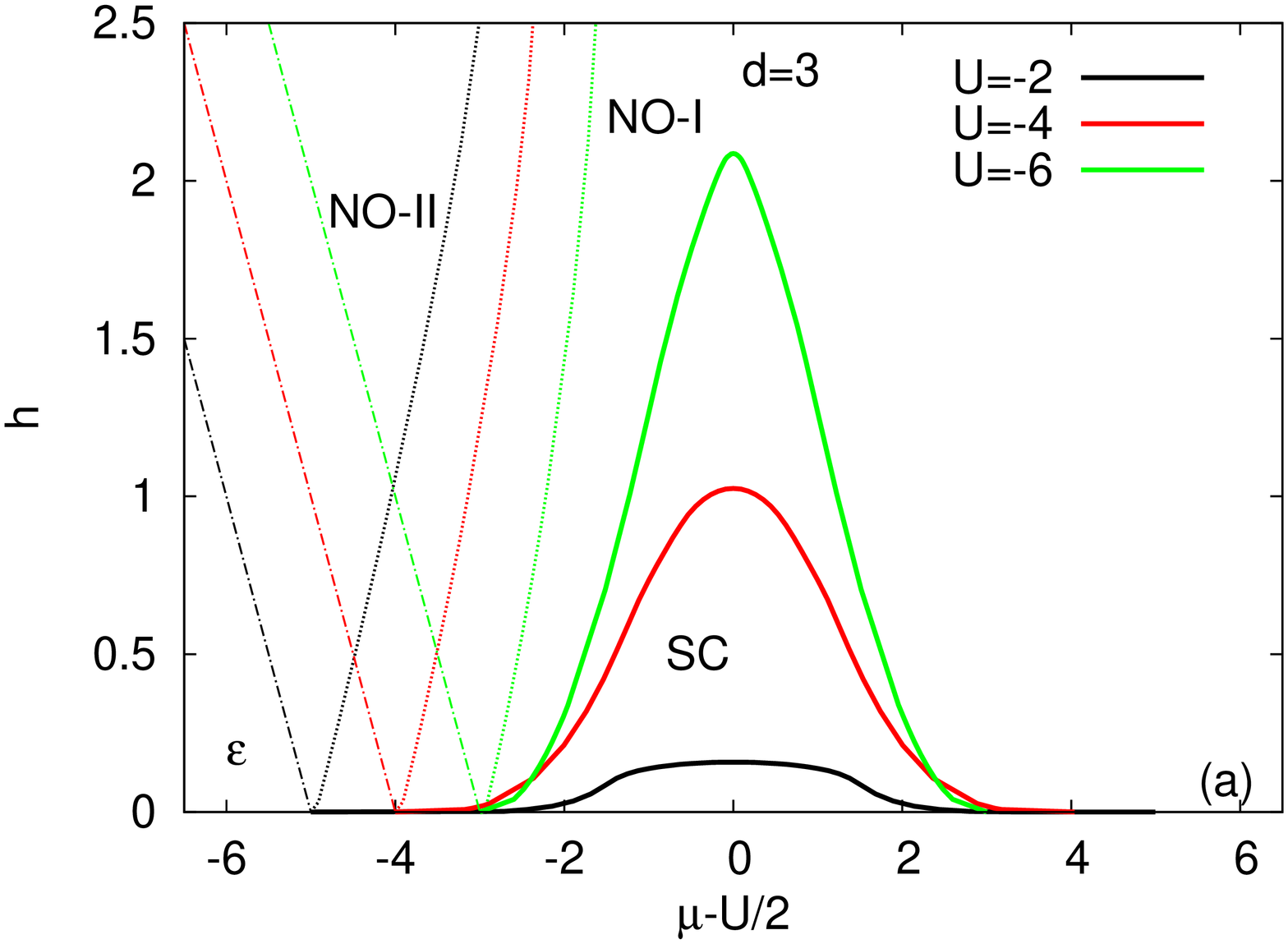}
\includegraphics[width=0.34\textwidth,angle=270]{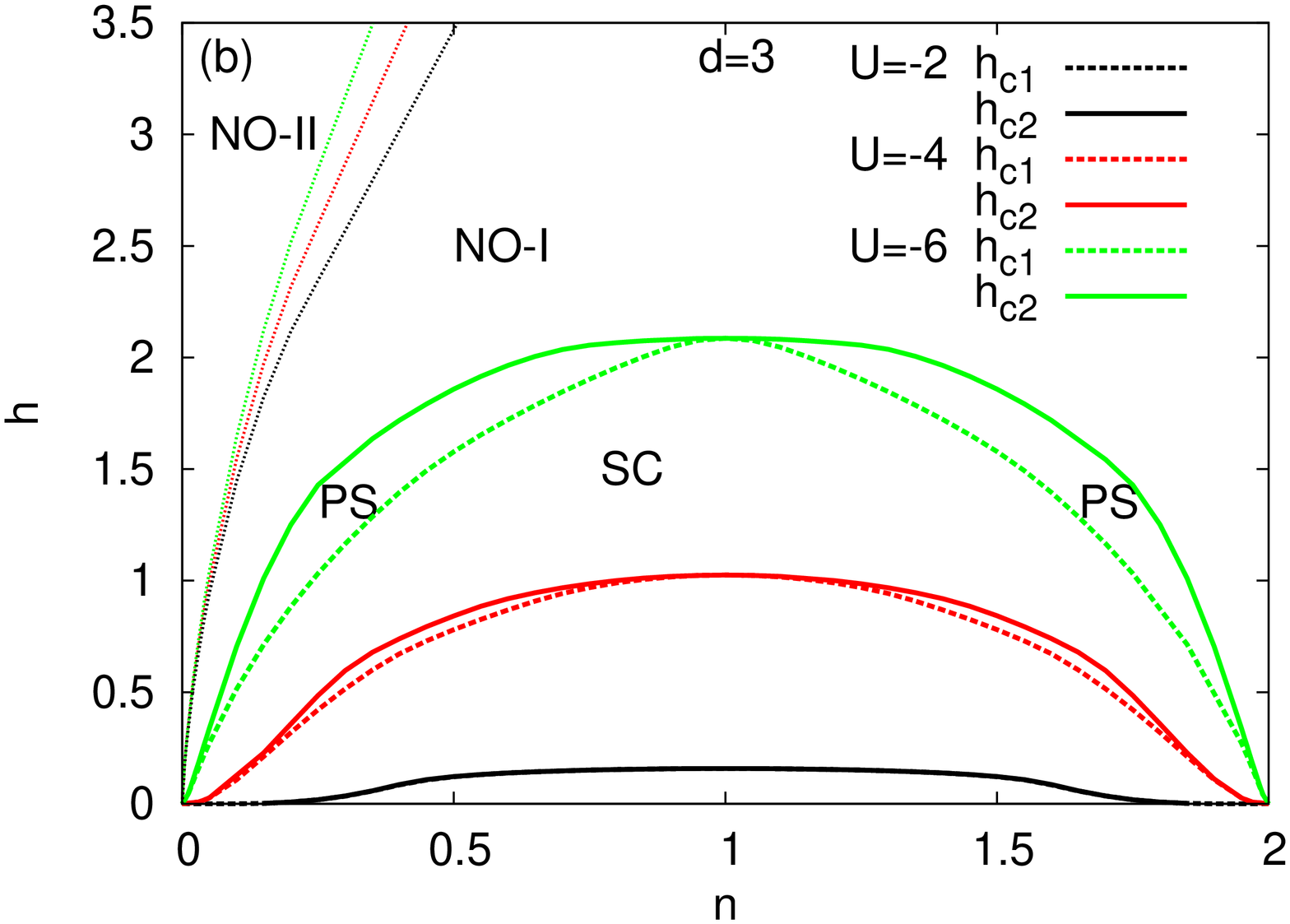}
\caption{The critical magnetic field vs. the chemical potential (a) and the electron concentration (b) for the first order SC-NO transition, at $T=0$; NO-I -- partially polarized ($P<1$) and NO-II -- fully polarized ($P=1$) normal state, $\varepsilon$ -- empty state, PS -- phase separation. In (a) solid lines denote SC-NO first order transition, dashed lines border NO-I and NO-II states, borders between NO-II and empty states are shown by dash-dotted line. Charge density wave (CDW) state, being degenerated with SC for h=0, n=1, is not shown.}
\end{center}
\end{figure}

 
Fig. 2 shows the magnetic field-temperature ($h-T$) phase diagram (3D case) for the fixed $n=0.75$ and $U=-4$. The lower set of curves denotes the diagram without the Hartree term. With increasing magnetic field, the character of the transition between SC and NO state changes from the second order (solid line) to the first order, which starts from the tricritical point (TCP). We can also observe the region of the phase separation between SC ($\Delta \neq 0$, $P\neq 0$) and NO ($\Delta =0$, $P\neq 0$) state in contrast to the ($T-h$) phase diagrams obtained for a fixed chemical potential \cite{kujawa}. The presence of the Hartree term raises the values of the critical magnetic field at $T=0$ for the first order transition, which exceed the Clogston limit. In the phase diagram without the Hartree term (conventional BCS like) the PS region is very narrow for the chosen values of the  parameters. For small $|U|$ and $n$ or for $n=1$, the Clogston limit is achieved in the phase diagrams without the Hartree term. For sufficiently high fields a reentrant transition takes place in the phase diagram with the Hartree term. Hence, the increase in temperature can induce superconductivity. We can infer that the Hartree term can be of importance in the context of the spin-polarized superconductivity and the neglect of this term may lead not only to quantitative but also to qualitative changes in the phase diagrams. We have also found the gapless superconducting (GSC) region (Fig. 2) i.e. spin-polarized SC, which has a gapless spectrum for the majority spin species for some magnetic fields ($h>\Delta$) and temperatures.

\begin{figure}[h!]
\begin{center}
\includegraphics[width=0.5\textwidth,angle=270]{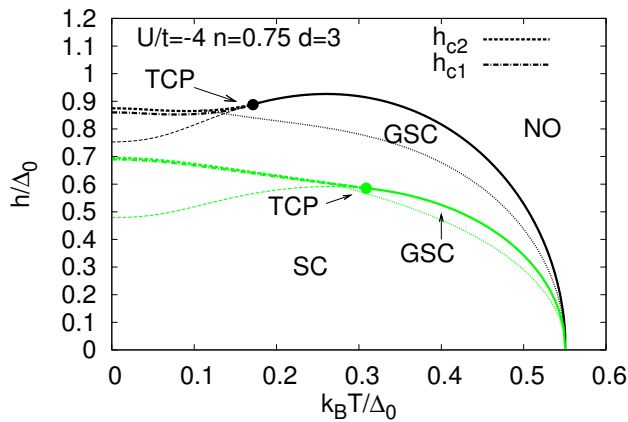}
\caption{Magnetic field-temperature phase diagrams for fixed $n=0.75$, $U=-4$, for a simple cubic lattice. The lower set of curves denotes the diagram without the Hartree term. The arrow indicates TCP. The thick dotted lines denote the first order phase transition. These lines bound PS region. GSC -- gapless superconductivity. The thin dotted line is merely an extension of the $2^{nd}$ order transition line below TCP (metastability line).}
\end{center}
\end{figure}

\section{Conclusions}
 We have considered the influence of a Zeeman magnetic field on superfluid properties of the attractive Hubbard model. The analysis has been restricted to the case of $s$-wave pairing with $\vec{q}=0$. For a fixed number of particles and $n\neq 1$, one obtains two critical magnetic fields in the phase diagrams, which limit the PS. 
In the PS region not only the particle densities but also the polarizations of the coexisting SC and NO states are different. We also observe a reentrant transition for sufficiently high magnetic fields in the $h-T$ phase diagram with the Hartree term. Moreover, the presence of the Hartree term causes an increase of $h_c$ of the Clogston limit. At $T>0$ the spin-polarized SC state and the gapless superconducting region have been found in $h-T$ phase diagrams.

\section*{Acknowledgements}
We would like to thank Prof. S. Robaszkiewicz for discussions.


\end{document}